\documentclass[nojss]{jss}
\usepackage{amsmath}
\usepackage{bm}
\usepackage{graphicx}

\author{
	Zachary Fisher\\
	\And 
	Elizabeth Tipton\\ 
}
\title{\pkg{robumeta}: An \proglang{R}-package for robust variance estimation in meta-analysis}

\Plainauthor{Zachary Fisher, Elizabeth Tipton} 
\Plaintitle{robumeta: an R package for robust variance estimation in  meta-analysis} 
\Shorttitle{Robust variance estimation in meta-analysis} 

\Abstract{
Meta-regression models are commonly used to synthesize and compare effect sizes. Unfortunately, traditional meta-regression methods are ill-equipped to handle the complex and often unknown correlations among non-independent effect sizes. Robust variance estimation (RVE) is a recently proposed meta-analytic method for dealing with dependent effect sizes. The \pkg{robumeta} package provides functions for performing robust variance meta-regression using both large and small sample RVE estimators under various weighting schemes. These methods are distribution free and provide valid point estimates, standard errors and hypothesis tests even when the degree and structure of dependence between effect sizes is unknown. 
}

\Plainkeywords{meta-analysis, meta-regression, robust variance estimation, small-samples} 

\begin{document}

\section{Introduction}

Meta-analysis refers to a collection of quantitative methods for combining evidence across studies \citep{hedges_statistical_1985}. Often this evidence is defined in terms of effect sizes, which are standardized so as to be comparable across studies. It is reasonable to treat these effect sizes as independent when estimators are derived from independent experiments or non-overlapping subject pools. However, non-independent effect sizes and effect size estimates are ubiquitous in published research \citep{van_three_2013} and ignoring or misspecifying their covariance structure can result in estimates that are not valid \citep{cooper_threats_2009}.  

Two common forms of dependence that arise in meta-analysis are correlated estimation errors and correlated effect size parameters.  Consider the simple model 

\begin{equation}
	T_{i}=\theta_{i}+\varepsilon_{i}
	\label{SimMod}
\end{equation}

where $T_{i}$ is an estimate of the $i^{th}$ effect size, $\theta_{i}$ is the true effect size, and $\varepsilon_{i}$ is the error resulting from our estimate of $\theta_{i}$ with $T_{i}$. When $\theta_{i}$ is fixed (or varies only as a function of known characteristics), dependence can occur through the estimation errors  ($\varepsilon_{i}s$). In practice, this form of dependence is common to studies where multiple estimates are obtained from the same units (e.g., multiple outcomes on the same individuals, multiple time points), or where a common control group is used for multiple experimental contrasts. 

When the $\theta_{i}s$ are considered random, the dependence can occur through the $\theta_{i}s$. Dependence among effect size parameters is commonly encountered when a hierarchical relationship exists among studies, such as reports published by the same laboratory, different groups using the same equipment, or multiple experiments reported in a single study. In this case, it is the studies which are nested within a larger category.  For a more detailed treatment of effect size dependency in meta-analysis see \citet{cooper_stochastically_2009} or  \citet{becker_multivariate_2000}. 

The problem with dependence is that traditional meta-analysis models assume independent effect sizes.  When studies report multiple effect sizes, two approaches are common.  One is to the use study average effect sizes, which result in a loss of information. A second approach is full multivariate meta-analysis \citep{hedges_statistical_1985, raudenbush_modeling_1988, kalaian_multivariate_1996}.  This is the optimal method for handling correlated data when the within-study covariance structure is known. Unfortunately, this information is rarely reported in published research. Without knowledge of the within-study correlations a number of factors should be considered before multivariate methods are used in practice. Those interested in a more detailed discussion of this topic should see \citet{jackson_multivariate_2011}. 

Recently,  \citet{hedges_robust_2010} provided a new approach for handling non-independent effect sizes that does not require knowledge of the within-study covariance. The RVE method extends work on heteroskedasticity-robust \citep{neyman_limit_1967, neyman_behavior_1967, white_heteroskedasticity-consistent_1980} and clustered \citep{liang_longitudinal_1986} standard errors in the general linear model, to the forms of dependence and weighting most often encountered in meta-analysis. The RVE approach has a number of desirable qualities. In meta-regression, the coefficients obtained from RVE are consistent estimates of the underlying population parameters under a broad set of conditions including non-normality. The results do not require the predictor variables to be fixed, as is the case in traditional regression methods used in meta-analysis\citep{hedges_fitting_1982}. Additionally, RVE produces valid standard errors, point estimates, confidence intervals, and significance tests when effect sizes are non-independent without needing to model the exact nature of this dependence. The RVE approach is increasingly being used in psychology, intervention research, medicine and ecology \citep[see][]{tipton_small_2013}. An overview of RVE methods geared towards practitioners, including a software tutorial for the \proglang{Stata} program \citep{hedberg_robumeta_2011} and \proglang{SPSS}, can be found in \citet{tannersmith_robust_2013}.

A previous limitation of RVE is that its application was limited to situations where the number of studies is large; this is because the results described in \citet{hedges_robust_2010} depend on asymptotic convergence.  Simulation studies on the rates of convergence for regression coefficients \citep{williams_using_2012}, the standardized mean difference \citep{hedges_robust_2010}, as well as the log odds ratio, log risk ratio, and risk difference effect sizes \citep{tipton_robust_2013}, suggested that while confidence interval coverage for intercepts was adequate with as few as 10 studies, the coverage rates for slope intervals were only nominal when the number of studies was greater than 40. This is problematic since work by \citet{ahn_review_2012} and \citet{polanin2013} suggests that more than 50\% of meta-analyses in psychology and education have fewer than 40 studies. 

More recently, \citet{tipton_small_2013} provided adjustments to the RVE estimators and degrees of freedom that greatly improve its performance in small samples.  These are particularly important when the number of studies is less than 40, but also when the covariates are unbalanced or highly skewed. \citet{tipton_small_2013} shows that it is hard to know a priori when these corrections are needed and suggest implementing them in all RVE analyses.  The \pkg{robumeta} package provides users with the ability to perform meta-regression with robust standard errors using both the original large-sample \citep{hedges_robust_2010} and small-sample adjusted \citep{tipton_small_2013} procedures. We briefly review these developments in RVE before proceeding to an example of RVE meta-regression using \pkg{robumeta}. 

\section{RVE}
Suppose we have $j=1,\ldots,m$ studies, each with with $k_{j}$ effect size estimates, $\mathbf{T}_{j}$. Let study $j$ also have a vector of $k_{j}$  residuals $\boldsymbol{\varepsilon}_{j}$, a design matrix $\mathbf{X}_{j}$, and weight matrix $\mathbf{W}_{j}$. The linear model relating the components can be written as 
\begin{equation}
	\mathbf{T}=\mathbf{X} \boldsymbol{\beta}+\boldsymbol{\varepsilon}
	\label{LinMod}
\end{equation}
where the vectors $\mathbf{T}=(\mathbf{T'}_{1},\ldots,\mathbf{T'}_{m})'$ and $\boldsymbol{\varepsilon}=(\boldsymbol{\varepsilon}'_{1}, \ldots \boldsymbol{\varepsilon}'_{m})'$ each contain $\mathrm{k}_{j}\:\mathrm{x}\:1$ stacked vectors corresponding to our $m$ studies, design matrix  $\mathbf{X}=(\mathbf{X}'_{1},\ldots,\mathbf{X}'_{m})'$ contains $m$ stacked $\mathrm{k}_{j}\:\mathrm{x\: p}$ matrices, and $\boldsymbol{\beta}=(\boldsymbol{\beta}'_{1},\ldots,\boldsymbol{\beta}_{\mathit{p}})'$ is a  $\mathrm{p}\:\mathrm{x\:1}$ vector of unknown regression coefficients.  For a given block diagonal matrix of weights  $\mathbf{W}=\mathrm{diag}(\mathbf{W}'_{1},\ldots,\mathbf{W}'_{m})'$, the weighted least squares estimates of $\boldsymbol{\beta}$ are 
\begin{equation}
	\mathbf{b}=\left(\sum_{j=1}^{m}\mathbf{X}'_{j}\mathbf{W}_{j}\mathbf{X}_{j}\right)^{-1}
	\left(\sum_{j=1}^{m}\mathbf{X}'_{j}\mathbf{W}_{j}\mathbf{T}_{j}\right).
	\label{beta}
\end{equation}
When the covariance of the $\boldsymbol{\varepsilon}_{j}$ in study $j$ is $\boldsymbol{\Sigma}_{j}$, the exact $\mathrm{p\:x\:p}$ covariance matrix for $b$ is

\begin{equation}
	\mathrm{V}(\mathbf{b)}=\left(\sum_{j=1}^{m}\mathbf{X}'_{j}\mathbf{W}_{j}\mathbf{X}_{j}\right)^{-1}\left(\sum_{j=1}^{m}	\mathbf{X}'_{j}\mathbf{W}_{j}\mathbf{\Sigma}_{j}\mathbf{W}_{j}\mathbf{X}_{j}\right)\left(\sum_{j=1}^{m}\mathbf{X}'_{j}		\mathbf{W}_{j}\mathbf{X}_{j}\right)^{-1}.
	\label{Vb}
\end{equation}

The problem that arises in estimating $\mathrm{V(\mathbf{b})}$ stems from difficulties in estimating the within-study covariance matrix $\boldsymbol{\Sigma}_{j}$. Although the individual variances are known, calculating the covariances of  $\boldsymbol{\varepsilon}_{j}$ in $\boldsymbol{\Sigma}_{j}$ depend on correlations that are often unreported. In RVE, the cross products of residuals for study $j$, $\boldsymbol{\mathrm{e}}_{j}\boldsymbol{\mathrm{e}}_{j}'$, are used as a rough estimate of the unknown $\boldsymbol{\Sigma}_{j}$. In doing so the estimate of  $\mathrm{V}(\mathbf{b})$ becomes 

\begin{equation}
	\mathbf{\mathrm{V}}^{R}=\left(\sum_{j=1}^{m}\mathbf{X}'_{j}\mathbf{W}_{j}\mathbf{X}_{j}\right)^{-1}\left(\sum_{j=1}^{m}			\mathbf{X}'_{j}\mathbf{W}_{j}\boldsymbol{\mathrm{e}}_{j}\boldsymbol{\mathrm{e}}_{j}'\mathbf{W}_{j}\mathbf{X}_{j}		\right)\left(\sum_{j=1}	^{m}\mathbf{X}'_{j}\mathbf{W}_{j}\mathbf{X}_{j}\right)^{-1}.
	\label{VR}
\end{equation}

Although the $\boldsymbol{\mathrm{e}}_{j}\boldsymbol{\mathrm{e}}_{j}'$ are rather poor estimates of the individual $\boldsymbol{\Sigma}_{j}$'s, the $\mathbf{\mathrm{V}}^{\mathrm{R}}$ estimator converges to the correct value, $\mathrm{V}(\mathbf{b)}$, as the number of studies in the meta-analysis $ m \rightarrow \infty$ \citep[see][Appendix]{hedges_robust_2010}. Inferences made on the regression coefficients are based on these robust standard errors.

\subsection{Weights}
In traditional meta-analysis, inverse variance weights play two roles, First, they produce the most efficient estimate of $\boldsymbol{\beta}$. Second, the statistical estimator of the variance requires the weights to be inverse; when they are not, the estimate of the variance can be severely biased. In RVE, inverse variance weights are also used, but here they are only important for efficiency. 

Weighting improves statistical efficiency by allocating more weight to studies that are more precise (i.e., have a smaller variance). Importantly, RVE provides asymptotically accurate estimates of the standard errors and valid inferential methods for any set of weights. This means the question of weights in RVE is only about efficiency. The most efficient weights would be $\mathbf{W}_{j}=\boldsymbol{\Sigma}_{j}^{-1}$. Clearly this is problematic since RVE is used when $\boldsymbol{\Sigma}_{j}^{-1}$ is unknown. For this reason, in RVE weights are developed based on a working model of the unknown covariance structure $\boldsymbol{\Sigma}_{aj}$ so that  $\mathbf{W}_{j}\approx\boldsymbol{\Sigma}_{aj}^{-1}$. Two working models (hierarchical and correlated effects) have been proposed to model the types of dependency most commonly encountered in meta-analytic research. Both of these weighting methods are only approximately efficient and in practice, researchers are urged to choose the weighting method based on the most common source of dependence in their data. 

\subsubsection{Correlated effects}

For study $j$, a working model based on the correlated effects dependence structure can be written as 

\begin{equation}
	\boldsymbol{\Sigma}_{aj}=\tau^{2}\mathbf{J}_{j}+\rho\nu_{j}(\mathbf{J}_{j}-\mathbf{I}_{j})+\nu_{j}\mathbf{I}_{j}
	\label{CEaj}
\end{equation}

where $\mathbf{J}_{j}$ is a $\mathrm{k}_{j}\:\mathrm{x\: k}_{j}$ all-ones matrix, $\mathbf{I}_{j}$ is a  $\mathrm{k}_{j}\:\mathrm{x\: k}_{j}$ identity matrix, $\nu_{j}$ is the estimation error variances for the $k_{j}$ effect sizes, and $\rho$ is a common correlation between the $k_{j}$ effect sizes. Since the correlated effects model assumes dependence arises from measurements made on the same number of subjects ($\varepsilon_{i}$), it is assumed the error variances $v_{ij}$ would be roughly constant within studies $\nu_{ij}=\nu_{j}$. Therefore, the simplified working model assumes a common correlation between within-study effect sizes. Approximate inverse variance weights are calculated by first partitioning the total study effect size variance equally among the  $k_{j}$ effect sizes, and then taking the inverse. More generally, we do not assume the error variances are constant within studies $\nu_{ij}\neq\nu_{j}$, and approximate inverse variance weights are calculated using 

\begin{equation}
	\mathbf{W}_{j}=w_{j}\mathbf{I}_{j}=\{1/[k_{j}(\nu_{\cdot j}+\tau^{2})]\}\mathbf{I}_{j}
	\label{Wj}
\end{equation}

where all the effect sizes in study $j$ are assigned the same weight $w_{ij}=w_{j}$, $\nu_{\cdot j}$ is the average study variance, and $\tau^{2}$ is an estimate of the between-study variance in study-average effect sizes.

To estimate $\tau^{2}$, we compute an initial meta-regression where all the effect sizes within study $j$ are given equal weights ($w_{j}=1/(k_{j}v_{j})$). Thus, the weighted residual sum of squares $\mathrm{Q_{E}}$ is

\begin{equation}
\mathrm{Q_{E}}=\sum_{j=1}^{m}\mathbf{T}'_{j}\mathbf{W}_{j}\mathbf{T}_{j} - \left(\sum_{j=1}^{m}\mathbf{T}'_{j}\mathbf{W}_{j}\mathbf{X}_{j}\right)\left(\sum_{j=1}^{m}\mathbf{X}'_{j}\mathbf{W}_{j}\mathbf{X}_{j}\right)^{-1} \left(\sum_{j=1}^{m}\mathbf{X}'_{j}\mathbf{W}_{j}\mathbf{T}_{j}\right),
\label{QE}
\end{equation}

and the method of moments estimate of between studies variance $\hat{\tau}^{2}$ is given by

\begin{equation}
	\hat{\tau}^{2}= 
	\frac
	{\mathrm{Q_{E}}-m+\mathrm{tr}
	\left[\mathbf{V}
	\left(\sum_{j=1}^{m}\frac{w_{j}}{k_{j}}\mathbf{X}'_{j}\mathbf{X}_{j}
	\right)\right]
	+ \rho \mathrm{tr}
	\left[\mathbf{V}
	\left(\sum_{j=1}^{m}\frac{w_j}{k_j}
	\left[\mathbf{X}'_{j}\mathbf{J}_{j}\mathbf{X}_{j} - \mathbf{X}'_{j}\mathbf{X}_{j}
	\right]\right)\right]}
	{\sum_{j=1}^{m}k_{j}w_{j} - \mathrm{tr} 
	\left[ \mathbf{V}
	\left(\sum_{j=1}^{m}w_{j}^{2}\mathbf{X}'_{j}\mathbf{J}_{j}\mathbf{X}_{j}
	\right)\right]
	}
	\label{Tau2}
\end{equation}
Note that this depends on an unknown value for the common correlation $\rho$. In practice, these results do not vary much for $\rho \in (0,1)$; later we will introduce a sensitivity approach that can be implemented in practice. 

\subsubsection{Hierarchical effects}

For study $j$, a working model based on the hierarchical dependence structure can be written as 

\begin{equation}
	\boldsymbol{\Sigma}_{aj}=\tau^{2}\mathbf{J}_{j}+\omega^{2}\mathbf{I}_{j}+\mathbf{V}_{j}
	\label{HEaj}
\end{equation}

where  $\mathbf{J}_{j}$ is a $\mathrm{k}_{j}\:\mathrm{x\: k}_{j}$ all-ones matrix, $\mathbf{I}_{j}$ is a $\mathrm{k}_{j}\:\mathrm{x\: k}_{j}$ identity matrix, $\mathbf{V}_{j}$ is a $\mathrm{k}_{j}\:\mathrm{x\: k}_{j}$ is a diagonal matrix of error variances for study $j$, $\omega^{2}$ is a measure of effect size variation within study $j$, and $\tau^{2}$ is a measure of effect size variation between studies. Approximate inverse variance weights for study j are retrieved from

\begin{equation}
	\mathbf{W}_{j}=\mathrm{diag}(w_{1j},\ldots,w_{kj})
	\label{wij}
\end{equation}

where $w_{ij}=1/(\nu_{ij}+\tau^{2}+\omega^{2})$. The weights are only constant in study $j$ if the error variances for each effect size are equal.

To estimate the parameters $\tau^{2}$ and $\omega^{2}$, we first obtain the weighted residual sum of squares using \eqref{QE}. We also calculate an additional sum of squares using the results from our preliminary meta-regression

\begin{equation}
    \mathrm{Q}_{1} =  \sum_{j=1}^{m} \left( \mathbf{T}_{j} - \mathbf{X}_{j} - \hat{\boldsymbol{\beta}} \right)' 
	\mathbf{J}_{j} \left( \mathbf{T}_{j} - \mathbf{X}_{j} - \hat{\boldsymbol{\beta}} \right).
	\label{Q1}
\end{equation}

\citet{hedges_robust_2010} provide the following method of moments estimators for $\tau^{2}$ and $\omega^{2}$

\begin{equation}
    \hat{\omega}_{2} = 
    \frac{A_{2} \left(\mathrm{Q}_{1}-C_{1} \right) - A_{1} \left( \mathrm{Q}_{E}-C_{2} \right)
    }
    {B_{1}A_{2}-B_{2}A_{1}
    }
	\label{Omega2}
\end{equation}

\begin{equation}
    \hat{\tau}_{2} = 
    \frac{\mathrm{Q}_{E}-C_{2}}{A_{2}} - 
    \hat{\omega}_{2} 
    \frac{B_{2}}{A_{2}}.
	\label{Tau22}
\end{equation}

Details on these equations can be found in \citet{hedges_robust_2010}.

\section{Small-sample adjustments}

As previously mentioned, while $\boldsymbol{\mathrm{e}}_{j}\boldsymbol{\mathrm{e}}_{j}'$ is a crude estimator of  $\boldsymbol{\Sigma}_{j}$ for individual studies, $\mathbf{V}^{R}$ converges in probability to $\mathbf{V}\mathrm{(\mathbf{b})}$ asymptotically \citep{hedges_robust_2010}. The properties, and limitations, of linearization estimators like $\mathbf{V}^{R}$ in small samples are well documented, and a number of solutions have been proposed to address these limitations. Interested readers can see  \citet{imbens_robust_2012} for a discussion of these solutions.

In this section, we review two types of small sample adjustments proposed by \citet{hedges_robust_2010} and extended by \citet{tipton_small_2013}. The first of these corrections is to the RVE estimator itself, while the second is to the degrees of freedom used for making inferences regarding the meta-regression coefficients $\boldsymbol{\beta}$. 

\subsection{Covariance structure adjustments}

The first adjustment is to the $\mathbf{V}^{R}$ estimator which can be written more generally as  

\begin{equation}
	\mathrm{\mathbf{V}}^{*}=\left(\sum_{j=1}^{m}\mathbf{X}'_{j}\mathbf{W}_{j}\mathbf{X}_{j}\right)^{-1}\left(\sum_{j=1}^{m}		\mathbf{X}'_{j}\mathbf{W}_{j}\mathbf{A}_{j}\boldsymbol{\mathrm{e}}_{j}\boldsymbol{\mathrm{e}}_{j}'\mathbf{A}_{j}\mathbf{W}_{j}		\mathbf{X}_{j}\right)\left(\sum_{j=1}^{m}\mathbf{X}'_{j}\mathbf{W}_{j}\mathbf{X}_{j}\right)^{-1},
	\label{Vstar}
\end{equation}

where in general $\mathbf{A}_{j}\mathbf{e}_{j}=\mathbf{A}_{j}(\mathbf{I}-\mathbf{H})_{j}\boldsymbol{\epsilon}$, $\mathbf{H}=\mathbf{X}\mathbf{Q}\mathbf{X}'\mathbf{W}$,  $\mathbf{Q}=(\mathbf{X}'\mathbf{WX})^{-1}$, and  $\boldsymbol{\epsilon}$ is a $\Sigma\mathrm{k}_{j}\:\mathrm{x\:1}$ vector of true residuals. \citet{hedges_robust_2010} proposed that this estimator could be improved by using 

\begin{equation}
\mathbf{A}_{j}^{HTJ}= [m/(m-p)]^{1/2}\mathbf{I}_{j}
\label{AjHTJ}
\end{equation}

where $\mathbf{I}_{j}$ is a $k_{j}$ by $k_{j}$ identity matrix. Simulations indicate that this adjustment is inadequate except in very specific cases and \citet{tipton_small_2013} develop a new solution based on extensions to the Bias Reduced Linearization (BRL) method proposed by \citet{mccaffrey_generalizations_2001}. Here since $\mathbf{V}^{*}$ also depends on $\mathbf{W}$ and working covariance matrices $\mathbf{\Sigma}_{aj}$, the $\Sigma\mathrm{k}_{j}$ by $\Sigma\mathrm{k}_{j}$ adjustment matrix $\mathbf{A}_{j}$ is specified separately for the hierarchical 

\begin{flalign}
	\qquad\qquad&\mathbf{A}_{j}^{H}=\mathbf{W}_{j}^{-1/2}[\mathbf{W}_{j}^{-1/2}(\mathbf{I}-\mathbf{H}_{jj})\mathbf{W}_{j}^{-3/2}]^{-1/2}\mathbf{W}_{j}^{-1/2}&
	\label{AJH}
\end{flalign}

and correlated effects weighting model

\begin{flalign}
	\qquad\qquad&\mathbf{A}_{j}^{C}=(\mathbf{I}-\mathbf{H}_{jj})^{-1/2},&
	\label{AJC}
\end{flalign}

where $\mathbf{H}_{jj} = \mathbf{X}_{j}\mathbf{Q} \mathbf{X}_{j}'\mathbf{W}_{j}$. Additionally, if non-efficient weights are required (in \pkg{robumeta} non-efficient weights can be specified using the \code{userweights} option) the adjustment matrix is specified as follows

\begin{flalign}
	\qquad\qquad&\mathbf{A}_{j}^{N}= \nu_{\cdot j}^{1/2} \left[\left(\mathbf{I}-\mathbf{H} \right)_{j}\mathbf{V}\left(\mathbf{I}-\mathbf{H} \right)_{j}'\right]^{-1/2}&
	\label{AJ}
\end{flalign}

where $(\mathbf{I}-\mathbf{H})_{j}$ includes the $k_{j}$ rows of $(\mathbf{I}-\mathbf{H})$ associated with study $j$ and the average variance of the study $j$'s effect sizes, $\nu_{\cdot j}=\left(1/k_{j}\right)\Sigma\nu_{ij}$.

In the above, for a general matrix $\mathbf{B}$ note that $\mathbf{I} = \mathbf{B}^{-1/2}\mathbf{BB}^{-1/2}$, and $\mathbf{B}^{-1/2} = \mathbf{P}\boldsymbol{\Lambda}^{-1/2}\mathbf{P}'$ where $ \mathbf{P}\boldsymbol{\Lambda}\mathbf{P}'$ is the eigen decomposition. See the Appendix of \citet{tipton_small_2013} for a detailed account of the development of these adjustments and their use with non-inverse variance weighting schemes.

\subsection{Degrees of freedom adjustment}

The second adjustment proposed by \citet{hedges_robust_2010} was to use the t-distribution with $df=m-p$ when making statistical inferences using $\mathbf{V}^{R}$. Building off of  \citet{bell_bias_2002}, however, \citet{tipton_small_2013} suggests using the Satterthwaite approximation \citep{satterthwaite_approximate_1946} to estimate the distribution of $\mathrm{\mathbf{V}}^{*}$ with $\chi^{2}_{df_{Sk}}$, where the  degrees of freedom $df_{Sk}$ is equal $2/c\nu^{2}$, and $c\nu$ is the coefficient of variation. 

These degrees of freedom are calculated separately for each regression coefficient under studying using 

\begin{equation}
df_{Sk}=(\sum\lambda_{jk})^{2}/\sum\lambda_{jk}^{2}
	\label{dfsk}
\end{equation}

where $\lambda_{jk}$ are the eigenvalues of $\mathbf{W}^{-1/2}(\sum \mathbf{g}_{jk}\mathbf{g}_{jk}')\mathbf{W}^{-1/2}$ where $\mathbf{g}_{jk}=(\mathbf{I}-\mathbf{H})_{j}'\mathbf{A}_{j}\mathbf{W}_{j}\mathbf{X}_{j}\mathbf{Q}l_{k}$. Here $l_{k}$ is a vector of $0$'s and $1$'s that indicates the regression coefficient, $\mathbf{A}_{j}$ and $\mathbf{W}_{j}$ are defined as before. Further details can be found in \citet{tipton_small_2013}. Simulations by  \citet{tipton_small_2013} show that the Satterthwaite approximation is valid so long as the $df>4$. This means that RVE can be effectively used to make inferences even with a very small number of studies. 

\section{Testing and confidence intervals}

Robust tests of the null hypothesis $H_{0}:\beta_{k}=0$ are conducted using the test statistic

\begin{equation}
t_{k}^{*}=b_{k}/\sqrt{\mathrm{V}_{kk}^{*}}
	\label{ttest}
\end{equation}

where $b_{k}$ is the estimate of $\beta_{k}$, $\mathrm{V}_{kk}^{*}$ is the $k^{th}$ diagonal of the $\mathbf{\mathrm{V}}^{*}$ matrix, and where we reject $H_{0}:\beta_{k}=0$ when $|t_{k}^{*}| \ge t_{df,{\alpha}}$. Additionally, a 95\% confidence interval for the $k$th regression coefficients $\beta_{k}$ is given by

\begin{equation}
b_{k}-t_{df, \alpha}\sqrt{\mathrm{V}_{kk}^{*}} \le \beta_{k} \le b_{k}+t_{df, \alpha}\sqrt{\mathrm{V}_{kk}^{*}}.
	\label{CI}
\end{equation}

Under the original \citet{hedges_robust_2010} approach, $\mathrm{V}^{*}_{kk}$ is calculated using equation \ref{VR} and $df=m-p$. When the small sample corrections from \citet{tipton_small_2013} are used, $\mathrm{V}^{*}_{kk}$ is calculated using $\mathbf{A}_{j}^{H}$, $\mathbf{A}_{j}^{C}$, $\mathbf{A}_{j}^{N}$ for the correlated effects, hierarchical effects and non-efficient weighting models, respectively. Here the degrees of freedom $df_{Sk}$, as specified in equation \ref{dfsk}, vary from covariate to covariate. 

\section[The robumeta package]{The \pkg{robumeta} package}

The \code{robu()} function is the main fitting function for \pkg{robumeta} and implements meta-regression using the large and small-sample RVE estimators described in \citet{hedges_robust_2010} and \citet{tipton_small_2013}, respectively. The arguments available in \code{robu()} are 

\begin{CodeInput}
robu(formula, data, studynum, var.eff.size, userweights, modelweights = 
     c("CORR", "HIER"), rho = 0.8, small = TRUE,...)
\end{CodeInput}

The \code{formula} object is similar to that found in other linear model specification routines (e.g., \code{lm()},  \code{glm()}). A typical formula will look similar to \code{y ~ x1 + x2...}, where \code{y} is a vector of effect sizes and \code{x1 + x2...} are the user-specified covariates.  An intercept only model can be specified with \code{y ~ 1} and alternatively, the intercept can be omitted by specifying \code{y ~ -1} and multiplicative interaction terms can be specified with \code{x1*x2...}. The \code{data} argument indicates the user-specified data frame which contains the following 3 columns

\begin{enumerate}
	\item Effect sizes (\code{effect.size}): User-specified column of effect sizes calculated from original research reports.
	\item  Effect size variances (\code{var.eff.size}): User-specified column of within-study variances for each effect size.
	\item  Study identification (\code{studynum}): User-specified column of study unique values used to identify studies. 
	\item  User-specified weights (\code{userweights}): User-specified weights if non-efficient weights are of interest (optional). 
\end{enumerate}
 
\code{modelweights} is the user-specified option for selecting either the hierarchical (\code{HIER}) or correlated (\code{CORR}) effects model. By default the correlated effects model is used. \code{rho} is the user-specified within-study effect size correlation, which must be specified when using the correlated effects model. The default value for \code{rho} is \code{0.8} and user-specified values must be between 0 and 1. \code{small = TRUE} or \code{small = FALSE} allow the user to specify whether or not the \citet{tipton_small_2013} adjusted estimation procedures will be used to fit the model,  \code{small = TRUE} is the default. The \code{robu()} function returns a fitted object of class \code{robu} which contains a list of summary information calculated during the model fitting procedure. The \code{print()} method provides a formatted output containing this summary information. 
 
\section{Examples}
\subsection{Example 1}

In the first example we use data from a meta-analysis conducted by \citet{oswald_predicting_2013} examining the predictive validity of the Implicit Association Test (IAT) and various explicit measures of bias for a variety of criterion measures of discrimination. The \code{oswald2013} dataset provided in \pkg{robumeta} contains $308$ effect sizes collected from $m=46$ individual studies. 

For this example, we will look at a subset of the \code{oswald2013.ex1} dataset which contains only the studies where the criterion measure of discrimination used was either neurological activity or response latency.  The (\code{oswald2013.ex1}) dataset contains $32$ effect sizes collected from $m=9$ individual studies. Additionally, the reported correlations have been transformed to  Fisher's $Z_{r}$ \citep{cooper_stochastically_2009}.  As the majority of effect-sizes in our dataset result from each study including multiple measures taken on the same individuals, we will use correlated effects weights to fit our model. 

\subsection{Intercept-only model}
When performing a meta-regression it is standard practice to first calculate an average effect size without conditioning on the study covariates. In RVE this can be followed with a sensitivity analysis which will be discussed in more detail in the following section. The fitting function for the intercept-only model is specified with the \code{robu()} function 

\begin{CodeInput}
oswald_intercept   <-  robu(formula = effect.size ~ 1, data = oswald2013.ex1, 
			studynum = Study, var.eff.size = var.eff.size, 
			rho = .8,  small = TRUE)
\end{CodeInput}

and the results displayed with \code{print()} as follows

\begin{CodeInput}
print(oswald_intercept)
\end{CodeInput}

\begin{CodeChunk}
\begin{CodeOutput}
RVE: Correlated Effects Model with Small-Sample Corrections 

Model: effect.size ~ 1 

Number of studies = 9 
Number of outcomes = 32 (min = 1 , mean = 3.56 , median = 2 , max = 11 )
Rho = 0.8 
I.sq = 82.05272 
Tau.Sq = 0.1849812 

            Estimate StdErr t-value   df P(|t|>) 95% CI.L 95% CI.U Sig
1 intercept    0.277  0.181    1.53 7.84   0.164   -0.141    0.695    
---
Signif. codes: < .01 *** < .05 ** < .10 *
---
Note: If df < 4, do not trust the results
\end{CodeOutput}
\end{CodeChunk}

In addition to the coefficient table, the output from \code{print()} contains the formula call used to fit the model, summary information for the total number of studies and outcomes included in the analysis, and a coefficient table. In addition, output for the correlated effects model contains the user-specified $\rho$ value, the estimate of between-study variance in study-average effect sizes $\tau^{2}$, and a descriptive statistic for the ratio of true heterogeneity to total variance across the observed effect sizes, $I^{2}$ \citep{higgins_measuring_2003}.   $I^{2}$ is calculated using 

\begin{equation}
I^{2} = [(\mathrm{Q_{E}} - df)/\mathrm{Q_{E}}]100\%
	\label{I.sq}
\end{equation}

where $\mathrm{Q_{E}}$ is defined as the weighted residual sum of squares as in equation \ref{QE}.  Unlike $\tau^{2}$, $I^{2}$ is not on the same scale as the effect size index used in the analysis and is better suited to anwsering questions regarding the proportion of observed variance resulting from real effect size differences. 

\subsection{Sensitivity analysis}

As mentioned previously, an estimate of $\tau^{2}$ is required to develop approximately efficient weights for the meta-regression model. The method of moments estimator for $\tau^{2}$ requires a value is specified for $\rho$.  For this reason \citet{hedges_robust_2010} suggests conducting a sensitivity analysis to determine the effect of $\rho$ on $\tau^{2}$. Generally, results from simulation studies suggest that $\tau^{2}$ and the meta-regression coefficients are relatively insensitive to changes in $\rho$  \citep{ishak_impact_2008, williams_using_2012, tipton_robust_2013}; however, if between-study covariances were to be considerably smaller than within-study covariances, these results might not hold. It should also be noted that a conservative approach and an external information approach have also been advocated for estimating $\rho$ in terms of developing efficient weights.  The conservative approach involves setting $\rho$ to $1$, which ensures studies don't receive additional weight due to having more effect sizes.  Additional information on these strategies can be found in \citet{hedges_robust_2010}.

The \code{sensitivty()} function in \pkg{robumeta} computes $\tau^{2}$, the average effect size and associated standard error for different values of $\rho$ in the interval $(0,1)$.

\begin{CodeInput}
R > sensitivity(oswald_intercept)
\end{CodeInput}

\begin{CodeChunk}
\begin{CodeOutput}
       Type  Variable rho=0 rho=0.2 rho=0.4 rho=0.6 rho=0.8 rho=1
1  Estimate intercept 0.277   0.277   0.277   0.277   0.277 0.277
2 Std. Err. intercept 0.181   0.181   0.181   0.181   0.181 0.181
3    Tau.Sq           0.184   0.184   0.184   0.185   0.185 0.185
\end{CodeOutput}
\end{CodeChunk}

From the sensitivity analysis it seems $\tau^{2}$ and subsequently the average effect size are relatively robust to different values of $\rho$. 

\subsection{Forest plot}

In meta-analysis, forest plots provide a graphical depiction of effect size estimates and their corresponding confidence intervals. The \code{forest.robu()} function in \pkg{robumeta} can be used to produce forest plots for RVE meta-analyses. The function requires the \pkg{grid} \citep{R} package and is based on examples provided in \citet{murrell_r_2011}. As is the case in traditional forest plots, point estimates of individual effect sizes are plotted as boxes with areas proportional to the weight assigned to that effect size.  Importantly, here the weight is not necessarily proportional to the effect size variance or confidence intervals, since the combined study weight is divided evenly across the study effect sizes. Two-sided 95\% confidence intervals are calculated for each effect size using a standard normal distribution and plotted along with each block. The overall effect is included at the bottom of the plot as a diamond with width equivalent to the confidence interval for the estimated effect.

\begin{figure}  
\centering
\includegraphics[width=5.3in]{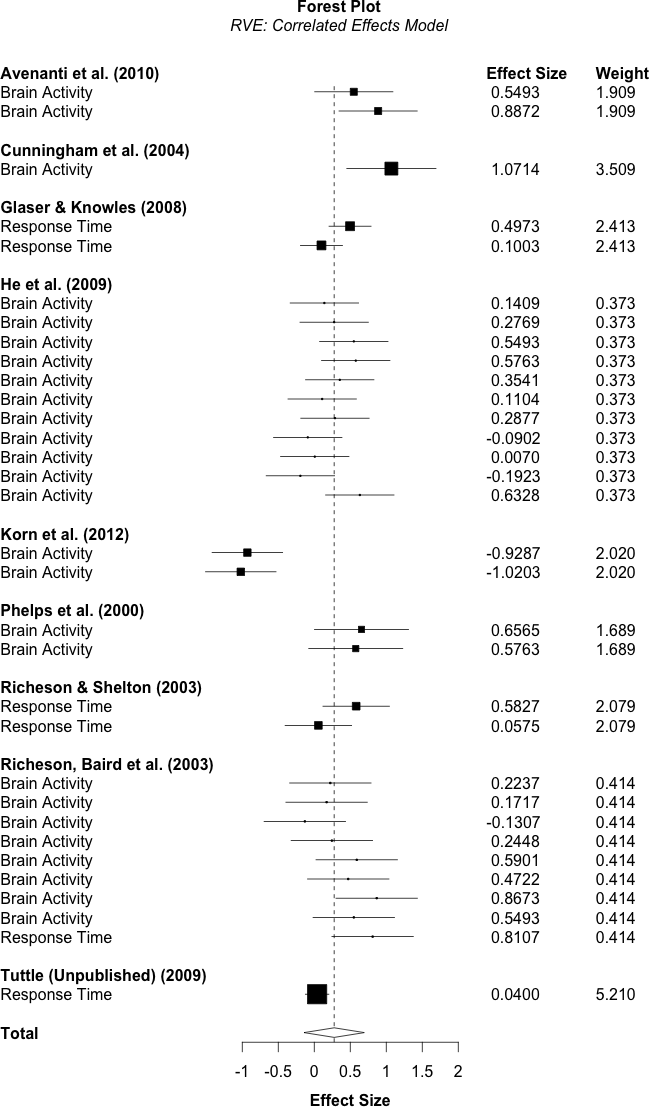}
\caption{\small \sl This figure shows a forest plot for the Correlated Effects model.}  
\label{fig:Forest}  
\end{figure}

The RVE forest function is designed to provide users with forest plots which display each individual effect size used in the meta-analysis, while taking into account the study- or cluster-level properties inherent to the RVE analysis. As such, the user must specify columns from their original dataset that contain labels for the study or cluster and for the individual effect sizes.  For instance, labels for the study column might be author names with corresponding publication years, and would be specified in the \code{forest.robu} function using the \code{study.lab} argument.  Labels for individual effect sizes might be ``Math Score'' or ``Reading Score'' for a meta-analysis that included such measures or as simple as ``Effect Size 1'' and ``Effect Size 2,'' and can be specified with the \code{es.lab} argument. Any number of additional columns can be specified to be plotted along side the confidence interval column and can be specified with the following syntax \code{"arg1" = "arg2"} where \code{"arg1"} is the title of the column on the forest plot, and \code{"arg2"} is the name of the column from the original \code{data.frame} that contains the information to be displayed.  An RVE forest plot with study labels corresponding to the author name and publication year, and effect size labels corresponding to the criterion for the Oswald data (see Figure \ref{fig:Forest} on page \pageref{fig:Forest}) is specified using the syntax below. In addition, a column of the RVE weights titled ``Weight'' and a column of effect sizes titled ``Effect Sizes'' is also included to illustrate the syntax for adding additional columns. 

\begin{CodeInput}
forest.robu(oswald_intercept, es.lab = "Crit.Cat", study.lab = "Study", 
            "Effect Size" = effect.size, # optional column
            "Weight" = r.weights)   # optional column
\end{CodeInput}

\subsection{Example 2}
In the following section we walk through an example where \pkg{robumeta} is used to fit simulated data described in \citet{tannersmith_robust_2013}. The dataset in question resulted from a fictional meta-analysis that examined the effectiveness of a brief alcohol intervention on alcohol intake. Intervention effectiveness was determined by differences in alcohol consumption between treatment and controls following the intervention. In this example standardized mean difference effect sizes are provided using Hedges' g, however, any effect size can be used with the \code{robu()} function. The \code{hierdat} dataset contains $m=15$ unique treatment centers with $68$ effect sizes. The covariates of interest in this example are \code{followup}, which indicates the length of the follow-up period before alcohol consumption was measured, in months and \code{binge} which indicates whether or not the alcohol consumption measure targeted binge drinking specifically. As the majority of dependency in this meta-analysis resulted from studies nested within treatment centers, we use the hierarchical effects weights to fit our model. 

\subsection{Covariate levels}

In RVE dependencies result from hierarchically structured data. Borrowing terminology from the multilevel and hierarchical linear modeling literature, this hierarchical structure takes the form of lower-level observations nested within higher-level groupings. Within the RVE framework we are generally concerned with two-level structures where level-1 (the lowest level) represents effect sizes in the correlated effects case and studies in the hierarchical effects case, while level-2 represents the study-level in the correlated effects case and a higher-order grouping level such as research lab in the hierarchical effects case. 

Since covariates can vary both between-level-2 and within-level-2 it is an important pre-processing step in any RVE analysis to identify the nature of this variation prior to model fitting.  For example, at first glance the \code{followup} variable appears to vary both within- and between-level-2. This indicates the length of follow-up periods varies both across studies conducted by the same treatment center, and then between the treatment centers themselves. Failure to distinguish between these effects can complicate interpretation later on as the regression coefficient actually represents a weighted combination of both between- and within-level-2 effects, which can be problematic when the effects differ in direction or magnitude. For instance, if treatment centers which produce higher quality studies also tend to employ longer follow-up periods the effect of follow-up length on treatment effect sizes might differ when examined within a given treatment center's studies, compared to the effect of follow-up length across treatment centers. 
 
To partition these two types of effect we can create group mean $X_{\bullet j}$ and group centered $X^{C}_{ij}$ versions of \code{followup} using the convenience functions \code{group.mean()} and \code{group.center()} provided in \pkg{robumeta}. 

\begin{CodeInput}
R > hierdat$followup_m <- group.mean(hierdat$followup, hierdat$studyid)
R > hierdat$followup_c <- group.center(hierdat$followup, hierdat$studyid)
\end{CodeInput}

Subsequently, including both \code{followup_m} and \code{followup_c} in our meta-regression model as follows

\begin{equation}
	T_{ij}=\beta_{0}+X_{\bullet j}\beta_{1}+X^{C}_{ij}\beta_{2}+\cdots
	\label{depReg}
\end{equation}

allows us to decompose the effects of follow-up length into two independent estimates where $\beta_{1}$ and $\beta_{2}$ represent the effect of a 1-month increase in $X_{\bullet j}$ and a 1-month increase in $X^{C}_{ij}$ on $T_{ij}$, respectively. Another important feature of centering is that it often indicates that a variable only varies at the study level, not effect size level.  We therefore encourage researchers to examine these variables before proceeding with their analysis. 

\subsection{Meta-regression model}
Now that we have completed the necessary data pre-processing we can fit our meta-regression model using hierarchical effects weighting with the following command  

\begin{CodeInput}
R > model_hier <- robu(effectsize ~ followup_c + followup_m + binge, 
			data = hierdat, modelweights = "HIER", studynum = 
			studyid, var.eff.size = var, small = TRUE)
\end{CodeInput}

The \code{print()} method can be used to present the results 

\begin{CodeInput}
R > print(model_hier)
\end{CodeInput}

\begin{CodeChunk}
\begin{CodeOutput}
RVE: Hierarchical Effects Model with Small-Sample Corrections 

Model: effectsize ~ followup_c + followup_m + binge 

Number of clusters = 15 
Number of outcomes = 68 (min = 1 , mean = 4.53 , median = 2 , max = 29 )
Omega.sq = 0.1650524 
Tau.Sq = 0.02479249 

              Estimate   StdErr t-value   df P(|t|>) 95% CI.L 95% CI.U Sig
1 intercept  -0.154226 0.147671  -1.044 6.07  0.3361 -0.51461  0.20615    
2 followup_c -0.000162 0.000695  -0.233 1.30  0.8470 -0.00534  0.00502    
3 followup_m  0.003467 0.002320   1.495 3.28  0.2243 -0.00357  0.01050    
4 binge       0.666645 0.115623   5.766 4.33  0.0035  0.35514  0.97815 ***
---
Signif. codes: < .01 *** < .05 ** < .10 *
---
Note: If df < 4, do not trust the results
\end{CodeOutput}
\end{CodeChunk}

The results of the meta-regression suggest the p-value for \code{followup_c} and \code{followup_m} are not reliable, as the Satterthwaite degrees of freedom are less than 4. This may result from the fact that center is contributing more than half of the effect sizes included in our meta-analysis did not actually vary the followup interval across studies, resulting in an unbalanced covariate. This is in line with results from simulation studies which suggest the Satterthwaite degrees of freedom are typically smaller for unbalanced covariates \citep{tipton_small_2013}. The effect of whether or not the study used a binge drinking measure (\code{binge}), however, was found to be positive and significant at $\alpha=0.05$. This suggests that studies which used binge drinking as their criterion measure were associated with increased treatment effect sizes compared to those studies that used some other measure of alcohol consumption.

\section{Conclusions}

Robust variance estimation (RVE) is a user friendly procedure grounded in work on \\heteroskedasticity-robust \citep{neyman_limit_1967, neyman_behavior_1967, white_heteroskedasticity-consistent_1980} and  clustered standard errors \citep{liang_longitudinal_1986}. RVE holds promise for alleviating the problems introduced by unknown within-study correlations in meta-analysis\citep{jackson_multivariate_2011}. This potential utility of RVE methods has been further expanded recently through the small- sample adjustments developed by \citep{tipton_small_2013} which greatly improve the finite-sample properties of RVE. The \pkg{robumeta} package provides functions for conducting meta-regression with both the large-sample \citep{hedges_robust_2010} and adjusted \citep{tipton_small_2013}  RVE estimators and provides functions for producing forest plots consistent with the RVE methodology. 

\bibliographystyle{jss}
\bibliography{bibtexrve}

\end{document}